# Magnetization and magneto-transport studies on $Fe_2VAl_{1-x}Si_x$

E P Amaladass[1,3], A T Satya[1], Shilpam Sharma[1], K Vinod[1], V Srinivas[2], C S Sundar[1] and A Bharathi[1,3]

[1] Condensed Matter Physics Division, Materials Science Group, Indira Gandhi Centre for Atomic Research, Kalpakkam-603102, India
[2] Department of Physics,
 Indian Institute of Technology Madras,
Chennai, 600036, India

[3] Corresponding Author: edward@igcar.gov.in
            : bharathi@igcar.gov.in

**Abstract:**
We report on magnetoresistance, Hall and magnetization measurements of $Fe_2VAl_{1-x}Si_x$ Heusler compounds for x= 0.005, 0.015, 0.02. There is a systematic change in the temperature coefficient of resistance (TCR) from negative to positive as the Si composition is increased. The Hall co-efficient shows that the carriers are electron like and the carrier density increases with Si concentration. Resistance measurements under magnetic field indicate a decreasing behavior under the application of magnetic field at low temperature region (T< 60 K), suggesting the suppression of scattering by magnetic field. Temperature and field dependent magnetization measurements did not show any significant change apart from the fact that the presence of super paramagnetic (SPM) cluster and its ordering at low temperatures. Arrott plot analysis of magnetization versus field also indicates the magnetic ordering with applied field below 60 K.

## 1. Introduction

$Fe_2VAl$ was thought to be a Kondo Insulator based on its non-magnetic ground state, with a high electronic mass from specific heat measurements [1]. Subsequent detailed studies revealed that the very small off-stoichiometry and anti-site disorder, gave this material a metallic ground state. Band structure calculations however, predicted a semi-metallic ground state, with a shallow density of states at Fermi level $E_F$. The calculations suggest that tuning the $E_F$ could result in the improvement of thermoelectric properties, since the density of states can be tinkered by changing $E_F$. Substituting Si for Al in $Fe_2VAl$ [2, 3] or substituting Al for Si in $Fe_2VSi$ [4] compound has been reported to show systematic change in electronic and thermal transport properties. For example, $Fe_2VAl$ alloy exhibits non-magnetic semiconductor like behavior while the iso-structural alloy $Fe_2VSi$ is found to be a semi metallic antiferromagnet (AFM) in spite of having similar band structure [5-7]. However, the AFM character of $Fe_2VSi$ could not be ascertained from the bulk susceptibility measurements [8, 9]. Band structure calculations predicted a ferromagnetic character in $Fe_2VAl_{0.5}Si_{0.5}$ alloy [10], but such a magnetic order was experimentally observed in 4 at% of *Si*-substituted $Fe_2VAl$ [2]. In this context it is interesting to see how the magnetic properties evolve upon Si substitution in $Fe_2VAl$ alloy across the insulator to metal transition. So far there is no systematic experimental study carried out in the literature. The puzzle includes the fundamental question whether the purported transport anomalies should be mainly attributed to the characteristic electronic band structure or structural/chemical disorder? Further it would be interesting to investigate whether the temperature co-efficient of resistivity gets affected on application of magnetic field? In this study, with the help of magnetization measurements we attempt to pin point, the nature of magnetic order that can lead to this effect. Temperature and field dependent resistivity and magnetization measurements have been carried out in samples, of $Fe_2VAl_{1-x}Si_x$ for x=0.005, 0.015 and 0.02.

## 2. Experiment

The alloy ingots of $Fe_2VAl_{1-x}Si_x$ (x=0.005, 0.015, 0.02) were prepared with high purity elemental constituents using an arc-melting under argon atmosphere. The as-melted ingots were sealed in evacuated quartz tubes and annealed at 1273 K for 48 hours and then quenched in cold water. Nominal composition assigned to each sample was regarded as accurate, because the weight loss was found to be less than 0.3 %. The electrical resistivity and magnetoresistance MR of samples were measured using a standard dc four terminal method in the temperature range of 3–300 K, using a home built cryostat with an accessible magnetic field upto 12 T. Hall effect measurements were done in a commercial (cryogenic, UK) cryofree, magnetoresistance system, operating in the 1.6 K to 300 K temperature range with accessible magnetic field upto 15 T. Magnetization measurements were performed using a commercial vibrating sample magnetometer (cryogenic, UK) in the temperature range of 4-300 K and magnetic field upto 16 T.

## 3. Results and discussion

### 3.1. *Temperature and field dependent resistivity*

Fig. 1 shows the temperature dependence of resistivity ($\rho$) of $Fe_2VAl_{1-x}Si_x$ for all three compositions of present study. As shown in the figure, a minute amount of Si substitution for *Al* causes a sharp decrease in electrical resistivity and the sign of the temperature coefficient of resistance (TCR) changes from negative for x=0.005 to a mixed TCR for x= 0.015 with minimum in the $\rho(T)$ curve. Eventually, the TCR becomes positive for x=0.02 suggesting the non-metal to metal transition with Si-Content. However, the changes in resistivity values either with concentration or with temperature do not appear to be that drastic to term the transition as Insulator-metal transition. Nevertheless, identification of semiconductor-metal transition, particularly in the vicinity of a transition is difficult. The decrease in the resistivity values along with change in TCR suggests non-metallic to metal state. In order to see the change in the carrier densities upon Si doping Hall Effect measurements were carried out at 4 K and essentially a linear variation of Hall resistance with field has been observed as shown in Fig. 2. Carrier concentration extracted assuming a single band model is shown in the inset of Fig. 2. It can be seen from the inset that the carrier concentration changes by a factor of two for a small change in Si concentration, and the TCR changes from negative to positive, for the same samples. The carrier concentration is ~$10^{20}$/cm$^3$, typically that of a degenerate semiconductor and the sign of the carrier is negative in agreement with earlier thermo-power data [10]. This fact can be taken as evidence that Si substituted Fe$_2$VAl is electron dominated compared to pure Fe$_2$VAl.

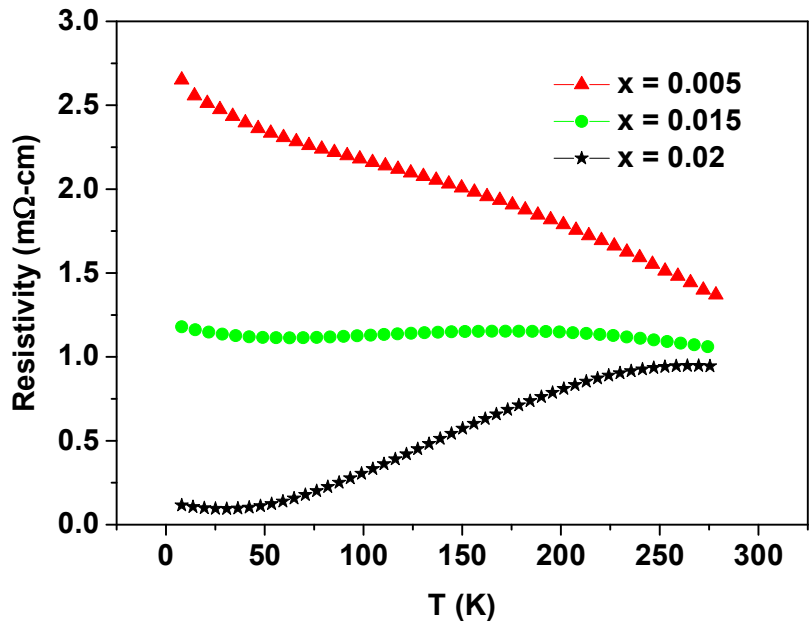

**Fig. 1.** Temperature dependence of resistivity for x= 0.005, 0.015, and 0.02.

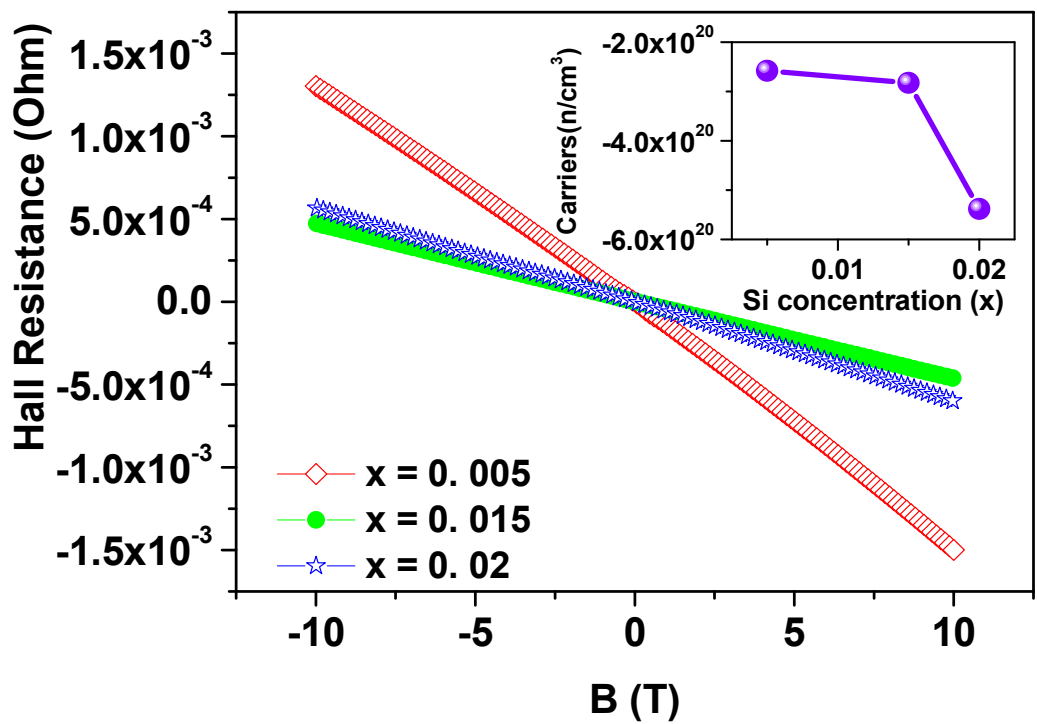

**Fig. 2.** Hall resistance versus magnetic field for x = 0.005, 0.015 and 0.02 at 4K. Inset: Carrier concentration as a function of Si fraction x.

In order to get insight into the origin of change in TCR, resistivity was measured at different magnetic fields in all the three samples and is plotted in the Fig. 3 (a)-(c). It is evident from the figure that the resistivity gets suppressed significantly at low temperatures upon application of magnetic field. In all the samples the change in the magnitude of resistance at the lowest temperature under the application of 10 T magnetic field, is of ~ 10% to ~2.5%. This taken in conjunction with the carrier concentration variation with Si substitution, could suggest that the percentage of carriers introduced by the substitution could be magnetic in nature in addition to the magnetism from anti-site defect present in these sample. Since magnetic impurities show Kondo scattering, we plot the resistivity versus logarithm of temperature. A clear linear behavior and its suppression in the presence of magnetic field further testifies that there is Kondo scattering in the system [11]. On close observation of the temperature variation of resistivity, $\rho(T)$, curves (Fig. 1) one can see a maximum in resistivity values around 200 K for x=0.015, which shift towards higher temperature on increasing Si content. These features suggest that at higher temperatures the ρ(T) may still exhibit non-metallic character.

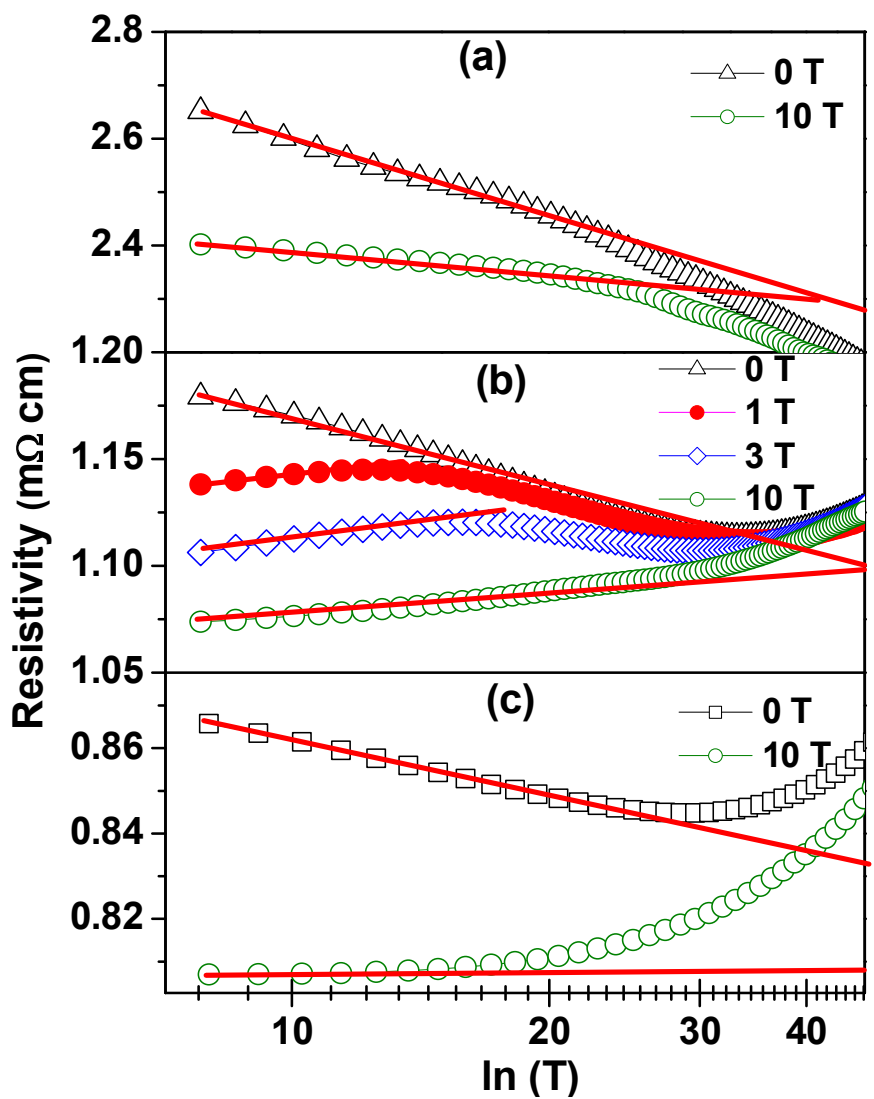

**Fig. 3.** Resistivity as a function of ln T (T ≤ 60 K) according to Kondo formalism for x=0.005 (a), 0.015 (b) and 0.02 (c). Good linear dependence shown by solid lines show that Kondo type scattering is dominant at low temperature and get suppressed upon applying magnetic field.

*3.2. Temperature dependence of Magnetization*

To elucidate the magnetic properties in the $Fe_2VAl_{1-x}Si_x$ (x=0-0.02) alloys, the temperature variation of magnetization for this system has been carried out through zero-field-cooled (ZFC) and field-cooled (FC) magnetization (M) curves under an external field of 0.01 and 1 T. As shown in the insets of Fig. 4 the FC/ZFC curves for x=0.005, 0.015 and 0.02 show a very distinct bifurcation between ZFC and FC curves from the room temperature. This strong irreversibility is a generic feature of disordered systems like cluster-glass, superparamagnet and inhomogeneous ferromagnet. Generally, such a bifurcation between ZFC and FC curves can be thought of as due to slow relaxation and competing interaction among interacting clusters in a magnetic system. However, from this point one cannot conclusively assert about the relevant magnetic phase of the system. ZFC and FC curves measured at 1 T (Fig. 4 (a-c)) do not show any bifurcation. Their features, such as, a broad maximum in the $M_{ZFC}(T)$ curve, the low temperature upturn, disappearance of irreversibility when the samples are cooled in higher applied field are suggestive of the presence of magnetic disorder in the present samples.

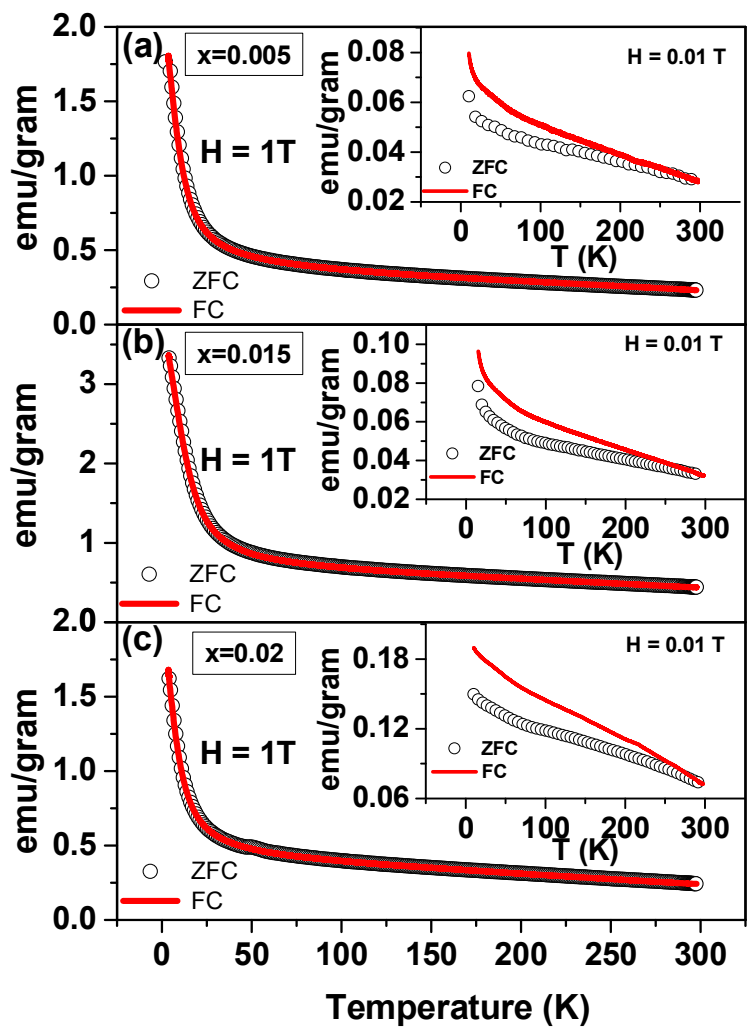

**Fig. 4**. Temperature variation of Magnetization [FC & ZFC] curves for $Fe_2VA_{1-x}Si_x$ (x=0.005, 0.015, 0.02) alloys at H = 1 T. Insets show the ZFC and FC curves of respective samples measured at H = 0.01 T.

*3.3. Field dependence of Magnetization*

To identify the nature of the magnetic state, isotherms have been recorded at specific temperatures (4 K, 10 K, 60 K, and 200 K) below and above the temperature $T_M$ where resistivity reaches maximum for x=0.015 sample. As observed in Fig. 5 (a-c) all compositions show similar trend in the M vs. H curve measured at different temperatures. At the lowest temperatures of measurement viz., 4 and 10 K the magnetization versus field curve show saturation behavior in fields as small as 2 to 3T, however no hysteresis was seen. This implies that no long range ferromagnetic order exists at low temperatures, but could arise from presence of magnetic clusters. For temperatures 60 K and 200 K a linear M versus H behavior are seen at high fields. The magnetization loops at 60 K and 200 K are fitted with a modified Langevin function [12, 13] that corresponds to the combination of super paramagnetic (SPM) and paramagnetic (PM) contribution as shown in equation.1.

$$M(H) = M_S [\coth(\alpha) - 1/\alpha] + \chi H \qquad (1)$$

$$\alpha = \mu H / K_b T$$

$M_S$ is the saturation magnetization and $\mu$ is the average magnetic moment of the magnetic clusters and $\chi$ is the susceptibility that account for the linear high field increase in the magnetization. Experimental data fit well to the model described above with fitting parameters shown in the Table. 1. The saturation magnetization, $M_S$ and the magnetic moment of SPM clusters, $\mu_{SPM}$ is found to be high for x=0.015 as compared to x=0.005 and 0.02. In addition the value of $\mu_{SPM}$ at 60 K is two orders of magnitude higher than $\mu_{SPM}$ at 200 K for all compositions. This shows that upon decreasing temperature (i) the interaction between isolated moments increases, (ii) the individual cluster size might be increasing, consequently the magnetization loops measured at 4 K and 10 K show saturation behavior. Absence of coercivity at low temperatures along with lower saturation magnetization values suggests presence of interacting magnetic clusters. Therefore in order to account for the temperature dependent cluster size variation and the inter-clusters interaction the experimental data is fitted with a modified Langevin function [13] as shown in equation below, which may be treated as two cluster model. It should be mentioned that the magnetization data fitted well to this model compared to the model with different combination of ferromagnet (FM), superparamagnet (SPM), and paramagnet (PM) viz., FM+PM, SPM+SPM, and SPM+PM.

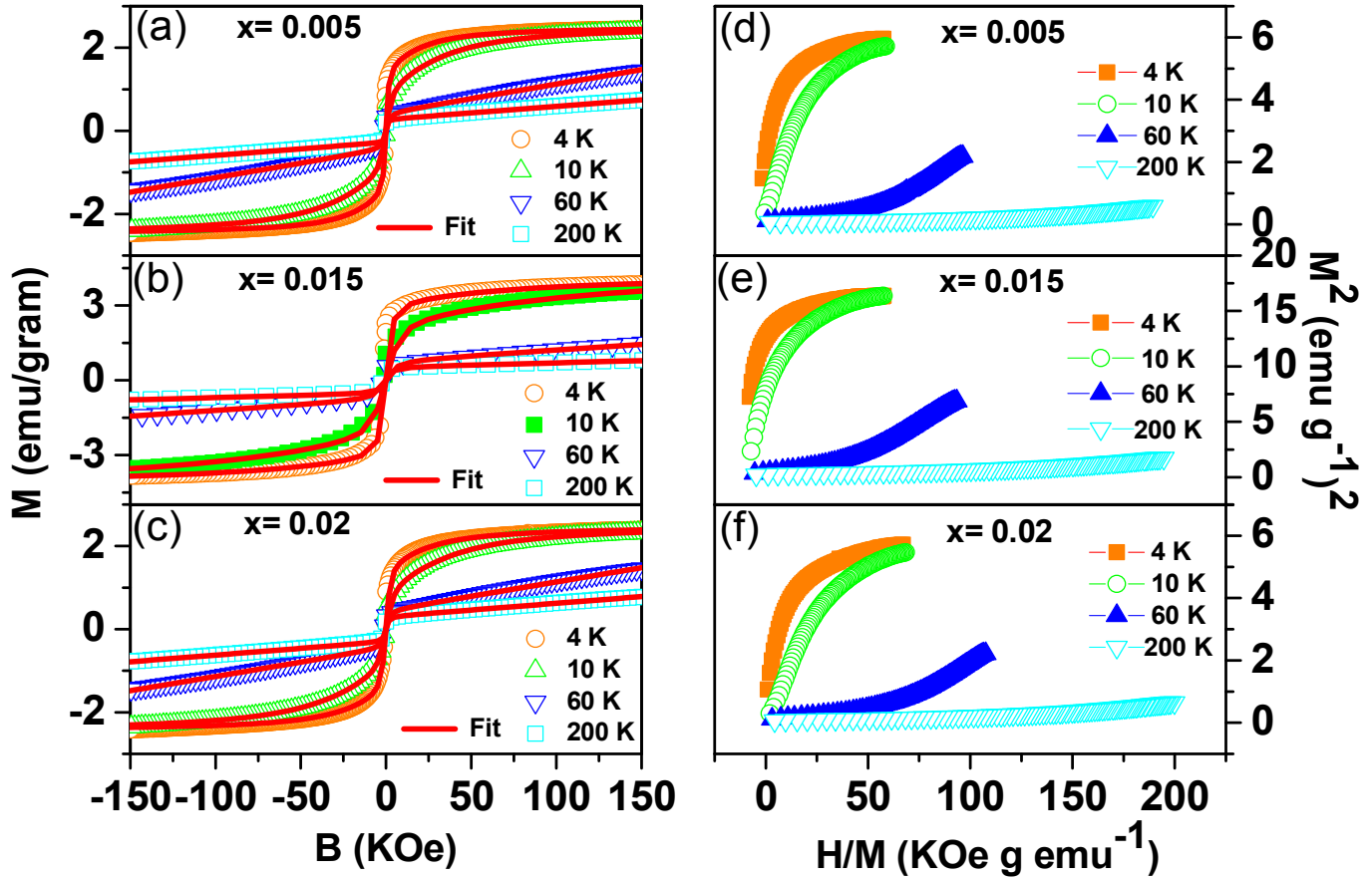

**Fig. 5.** Magnetic field dependent magnetization at different temperatures (4 K, 10 K, 60 K, 200 K) for x= 0.005 (a), x=0.015 (b), and x=0.02 (c). Solid lines are fit to the experimental data using modified Langevin function. (d)(e)(f) Arrott plot ($M^2$ vs. ($H/M$)) for respective composition.

$$(M(H)-M_0)/M_s = f * [\text{Coth}(\alpha_1) - 1/\alpha_1] + (1-f) * [\text{Coth}(\alpha_2) - 1/\alpha_2] \qquad (2)$$

$\alpha_1 = \mu_1 H/K_b T$

$\alpha_2 = \mu_2 H/K_b T$

Here $\mu_1$ and $\mu_2$ corresponds to the cluster magnetic moment (cluter1, cluster2) and *f* its respective fraction. $M_0$ corresponds to the field independent magnetization and $M_s$ is saturation magnetization.

**Table. 1** Fitting parameters extracted from fitting the magnetization vs. magnetic field data at 60 K and 200 according to equation.1.

| composition x | 60 K | | | 200 K | | |
|---|---|---|---|---|---|---|
| | $M_s$ (emu/g) | $\mu_{SPM}$ ($10^2 \mu_B$) | $\chi$ (emu/g T) | $M_s$ (emu/g) | $\mu_{SPM}$ ($\mu_B$) | $\chi$ (emu/g T) |
| 0.005 | 0.436 | 10.87 | 0.069 | 0.0276 | 9.35 | 0.031 |
| 0.015 | 0.789 | 11.32 | 0.131 | 0.537 | 15.02 | 0.050 |
| 0.02 | 0.458 | 10.03 | 0.068 | 0.303 | 9.56 | 0.032 |

The fitting parameter obtained after fitting the data are shown in Table 2. At 4 K the sample with x=0.015 shows the large fraction of cluster1 with large magnetic moment as compared to other compositions. Further increasing the composition the magnetic moment and its fraction decreases. Though the data at 4 K and 10 K fit well to the two-cluster model, the cluster moment shows an increasing behavior upon increasing temperatures. Whereas the clusters moment obtained by SPM + PM contribution (equation.1 for 60 K and 200 K) decreases with temperature and very large cluster moment is seen at 60 K.

**Table. 2** Fitting parameters extracted from fitting the magnetization data at 4 K and 10 according to two cluster model in equation.2.

| composition x | 4 K | | | | 10 K | | | |
|---|---|---|---|---|---|---|---|---|
| | $M_s$ (emu/g) | f % | $\mu_1$ ($\mu_B$) | $\mu_2$ ($\mu_B$) | $M_s$ (emu/g) | f % | $\mu_1$ ($\mu_B$) | $\mu_2$ ($\mu_B$) |
| 0.005 | 2.50 | 63.27 | 104.37 | 5.10 | 2.56 | 41.51 | 111.08 | 8.05 |
| 0.015 | 4.15 | 77.62 | 143.64 | 4.14 | 4.30 | 57.56 | 144.99 | 7.58 |
| 0.02 | 2.47 | 65.76 | 61.18 | 3.89 | 2.53 | 41.15 | 90.32 | 7.10 |

Further the Arrot plots for all three compositions are shown in Fig. 5 (d-f). Mean field theory predicts that the H/M plot should be linear in $M^2$. Below Curie temperature $T_C$ the curves intercept the $M^2$ axis and provide the information on spontaneous magnetization. At $T_C$ the line passes through origin and for $T > T_C$ it intercept the H/M axis [14]. Nonlinear behavior seen in Fig. 5 suggests the absence of FM to PM transition in the system. For 4 K and 10 K the curves show a strong curvature towards H/M axis whereas for 60 K and 200 K an opposite curvature is seen. The spontaneous magnetization values extracted by forced extrapolation of Arrot plots provide small values. These results suggest the non-interacting SPM cluster and weak magnetic behavior in the sample. The change in curvature of the Arrot plot for 10 K ≤ $T_o$ ≤ 60 K suggest that at $T_o$ the system undergoes field induced ordering of the randomly oriented SPM clusters. These observations are consistent with the magnetic isotherms analysis discussed earlier in this work. Further low magnetic moments could be due to the development of antiferromagnetic interactions on Si-substitution. Such clusters can also generate large magnetoresistance values.

**Table. 3** Comparison of resistivity at 300 K and magnetization value at 5 K measured at 5 T.

| Sample $Fe_2VAl_{1-x}Si_x$ | Resistivity at 300 K | Magnetization at 5 K and 5T |
|---|---|---|
| x=0 (as casted) [12] | 1.35 mOhm | 2.70 emu/g |
| x=0 (annealed) [12] | 1.51 mOhm | 4.23 emu/g |
| x=0.005 | 1.40 mOhm | 2.19 emu/g |
| x=0.015 | 1.10 mOhm | 3.75 emu/g |
| x=0.02 | 0.90 mOhm | 2.09 emu/g |

The ordering of SPM clusters is further clarified by comparing the magnetization and the resistivity data. Temperature dependent magnetization data shows a saturation behavior below 60 K whereas resistivity data shows a bifurcation of resistivity ($\rho_{min}$) at T ≤ 60 K. Therefore 60 K is believed to be the weak ordering temperature of the SPM species in the sample as also inferred from the Arrot plot. In the absence of field, as the temperature decreases the interaction among these SPM clusters increases and results in high resistivity due to strong scattering of carriers. In high magnetic fields (10 T) the SPM clusters gets aligned with the fields and consequently the resistance is decreased in the sample.

*3.4. Resistance versus magnetic field*

The normalized resistivity versus magnetic field measured at 10 K is shown in Fig. 6. It is evident from the figure that the resistivity change in the presence of magnetic field is higher in the lower concentration samples viz., x=0.005 and x=0.015. The observed lower magneto-resistance in the x=0.02 sample is surprise since the carrier concentrations in the x=0.015 and x=0.02 samples are different just by a factor of two. This could imply that magnetic moment/ magnetic ordering are qualitatively different in the two samples; this could arise due to sample dependent anti-site disorder, which could have a different magnetic nature in the two samples. Notably in the x=0.02 sample the low temperature saturation magnetization is much lower than that seen in the x=0.015 sample. Previous reports Ref. [12] showed that the annealed $Fe_2VAl$ sample show an increase in magnetization and resistivity due to less crystal defects after annealing. Table.3 compares the value of magnetization and resistivity of the pristine sample from Ref. [12] with that of the Si doped samples. Resistivity shows a systematic decrease upon Si doping whereas values of magnetization show non monotonous behavior. This shows that the anomalous behavior is due to sample dependent anti-site defect states or due to emergence of new electronic states from the interaction of Si with Fe and V as reported for $Fe_2VAl_{1-x}Ge_x$ alloy [15].

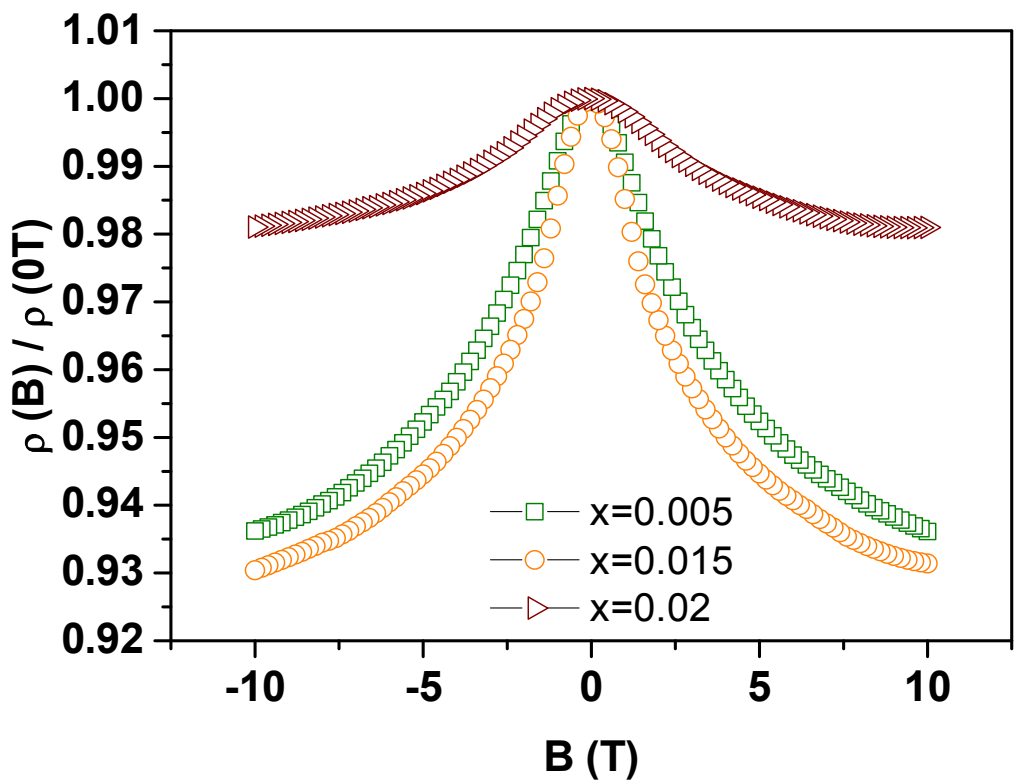

**Fig. 6.** Normalized magnetoresistance plots at 10 K for all compositions (x = 0.005, 0.015, 0.02)

## 4. Conclusion

A systematic study of the magnetization and magneto transport properties of $Fe_2VAl_{1-x}Si_x$ was performed. Upon very minute substitution of Si (x = 0.005, 0.015, 0.02) for Al, the $Fe_2VAl_{1-x}Si_x$ alloy shows a change in the sign of TCR. Kondo type magnetic scattering at low temperature was present for all compositions. The crossover from negative to positive TCR takes place at x = 0.02. Intermediate composition, x= 0.015 shows a mixed TCR behavior and high magnetoresistance. The hall data shows systematic increases in the carrier concentration upon Si substitution. Temperature and field dependent magnetization did not show any significant change apart from the fact that the presence of SPM cluster in the sample and its field induced FM ordering at low temperatures. Being a narrow gap semiconductor with localized density of states from anti-site defects, very minute increase in composition (x= 0.005, 0.015, 0.02) changes the density of the charge carrier and results in the transition from negative to positive TCR. Magnetic state of the sample remain less perturbed apart from the increase in magnetization and magnetoresistance for intermediate composition, x= 0.015 which might be due to sample dependent anti-site defects.

**Acknowledgement**

E.P.A acknowledges the financial grant as KSKRA fellowship from Board of Research in Nuclear Sciences (BRNS), BARC, Trombay, Mumbai, India is also acknowledged.